\documentclass[%
reprint,
superscriptaddress,
nofootinbib,
bibnotes,
amsmath,amssymb,
aps,
]{revtex4-1}

\usepackage{graphicx}% Include figure files
\usepackage{dcolumn}% Align table columns on decimal point
\usepackage{bm}% bold math
\usepackage{hyperref}% add hypertext capabilities
\usepackage{graphicx,slashed,hyperref,color,subfig}
\usepackage{amssymb}
\usepackage[normalem]{ulem}
\usepackage[utf8]{inputenc}
\hypersetup{
   bookmarks=true,         % show bookmarks bar?
   unicode=true,          % non-Latin characters in Acrobat�s bookmarks
   pdftoolbar=true,        % show Acrobat�s toolbar?
   pdfmenubar=true,        % show Acrobat�s menu?
   pdffitwindow=false,     % window fit to page when opened
   pdfstartview={FitH},    % fits the width of the page to the window
   pdftitle={My title},    % title
   pdfauthor={Author},     % author
   pdfsubject={Subject},   % subject of the document
   pdfcreator={Creator},   % creator of the document
   pdfproducer={Producer}, % producer of the document
   pdfkeywords={keyword1} {key2} {key3}, % list of keywords
   pdfnewwindow=true,      % links in new PDF window
   colorlinks=true,       % false: boxed links; true: colored links
   linkcolor=blue,          % color of internal links (change box color with linkbordercolor)
   citecolor=magenta,        % color of links to bibliography
   filecolor=magenta,      % color of file links
   urlcolor=cyan           % color of external links
}

\def\lsim{\mathrel{\rlap{\lower4pt\hbox{\hskip1pt$\sim$}}
    \raise1pt\hbox{$<$}}}         %less than or approx. symbol
\def\gsim{\mathrel{\rlap{\lower4pt\hbox{\hskip1pt$\sim$}}
    \raise1pt\hbox{$>$}}}         %greater than or approx. symbol

\begin{document}

\preprint{LDU2018-2}

\title{Minimal gauge inflation and the refined Swampland conjecture}% Force line breaks with \\
%\thanks{This work was supported in part by the National Research Foundation of Korea (NRF) grant funded by the Korean government (MSIP) (No. 2016R1A2B2016112) and (NRF-2018R1A4A1025334)
%}%
\author{Seong Chan Park}
\email{sc.park@yonsei.ac.kr}
\affiliation{Department of Physics \& IPAP, Yonsei University, Seoul 03722 Korea}

%% \altaffiliation[Also at ]{Physics Department, XYZ University.}%Lines break automatically or can be forced with \\

\date{\today}% It is always \today, today,
             %  but any date may be explicitly specified

\begin{abstract}
The refined de Sitter (dS) conjecture provides two consistency conditions for an effective theory potential of a quantum gravity theory. Any inflationary model can be checked by these conditions and minimal gauge inflation is not an exception. We develop a generic method to analyze a monotonically growing potential with an inflection point on the way to the plateau near the top such as the potential  in minimal gauge inflation model and the Higgs inflation.  Taking the latest observational data into account, we find the fully consistent parameter space where the model resides in the Landscape rather than in the Swampland.

\end{abstract}

\pacs{Valid PACS appear here}% PACS, the Physics and Astronomy
                             % Classification Scheme.
%\keywords{Suggested keywords}%Use showkeys class option if keyword
                              %display desired
\maketitle

%%%%%%%%%%%%%%%%%%%%%%%%%%%%%%%%%%%%%%
\section{ Introduction}
\label{sec:introduction}
%%%%%%%%%%%%%%%%%%%%%%%%%%%%%%%%%%%%%%

The refined de Sitter (dS) conjecture, the latest version of the Swampland conjecture, was recently proposed by Ooguri, Palti, Shiu and Vafa~\cite{Ooguri:2018wrx} after an earlier version proposed by Obied, Ooguri, Spondyneikeo and Vafa~\cite{Obied:2018sgi}. The conjecture states that any scalar potential $V(\phi)$ for scalar fields in a low energy effective theory of a consistent  quantum gravity must satisfy at least one of the following conditions:
\begin{align*}
&||\nabla V|| \geq c_1 \frac{V}{M_P},&\text{(Condition-1)}\\
%&~~~~~~~~~~~~~~~~~~~~~~~~~~~\text{or}\\
&{\rm min}(\nabla_i \nabla_j V) \leq -c_2 \frac{V}{M_P^2},&\text{(Condition-2)}
\end{align*}
where $c_1$ and $c_2$ are universal, positive constants of  order unity and ${\rm min}(\nabla_i \nabla_j V)$ of the second condition is the minimum eigenvalue of the Hessian $\nabla_i \nabla_j V$ in an orthonormal frame.\footnote{In literatures, $c$ and $c'$ are also widely used instead of $c_1$ and $c_2$ in this paper.
Note $c_1$ and $c_2$ are for the first and the second derivatives of the potential, respectively.}

One of the most straightforward and intriguing implications of the conjecture is that the cosmological constant scenario, for which $||\nabla V_{c.c}||=0$ and $V_{c.c.}>0$, is ruled out but the quintessence field with an exponentially decaying potential $V_{Q} \left(Q\right) = \Lambda_{Q}^{4} e^{-c_{Q} Q}$ would consistently explain the late time expansion if $c_Q \geq c_1$~(see a recent review of quintessence model~\cite{Tsujikawa:2013fta}).  There have been many follow-up papers considering various implications of the conjecture~\cite{Fukuda:2018haz,Wang:2018kly, Das:2018rpg, Antoniadis:2018ngr, Ashoorioon:2018sqb, Motaharfar:2018zyb, Odintsov:2018zai,Dimopoulos:2018upl,Kawasaki:2018daf,Hamaguchi:2018vtv, Lin:2018kjm, Anguelova:2018vyr, Ellis:2018xdr, Halverson:2018cio, Brandenberger:2018xnf, Bena:2018fqc, Moritz:2018ani, Visinelli:2018utg, DAmico:2018mnx, Han:2018yrk, Brandenberger:2018wbg}.

Distinctively from the (Condition-1), the newly added (Condition-2) is rather easily satisfied for a generic potential in low scale, $\Delta \phi \ll M_P$,  because
\begin{eqnarray}
M_P^2 \frac{\nabla_i \nabla_j V}{V} \sim - \frac{M_P^2}{\Delta \phi^2} \ll -c_2 \sim -\mathcal{O}(1).
\end{eqnarray}
However, inflationary dynamics, which takes place at a high scale, is constrained by the conditions  as analyzed 
%in this light for natural inflation, pure natural inflation, Starobinsky model and $\alpha$-attractor models 
recently in Ref.~\cite{Fukuda:2018haz}. The dS conjecture indeed provides some new insights in viewing each model so that  we are motivated to examine a new kind of model, which has not been examined so far: a minimal gauge inflation model introduced in Ref.~\cite{Gong:2018jer}. 

In the next section, Sec.~\ref{sec:model}, we start from the inflaton potential of minimal gauge inflation model and apply the dS conjecture to read out the consistency conditions in generic field space.  In Sec.~\ref{sec:predictions}, now considering the latest cosmological observations from the Planck 2018 and also the polarization measurements from BICEP/{\it Keck}, we re-examine the dS conditions for a fully realistic case, which can provide the most interesting understanding of the model in light of the dS conjecture.  Finally we conclude in Sec.~\ref{sec:conclusion}.

\begin{figure}[t]
\centering % \begin{center}/\end{center} takes some additional vertical space
\includegraphics[width=.88\columnwidth]{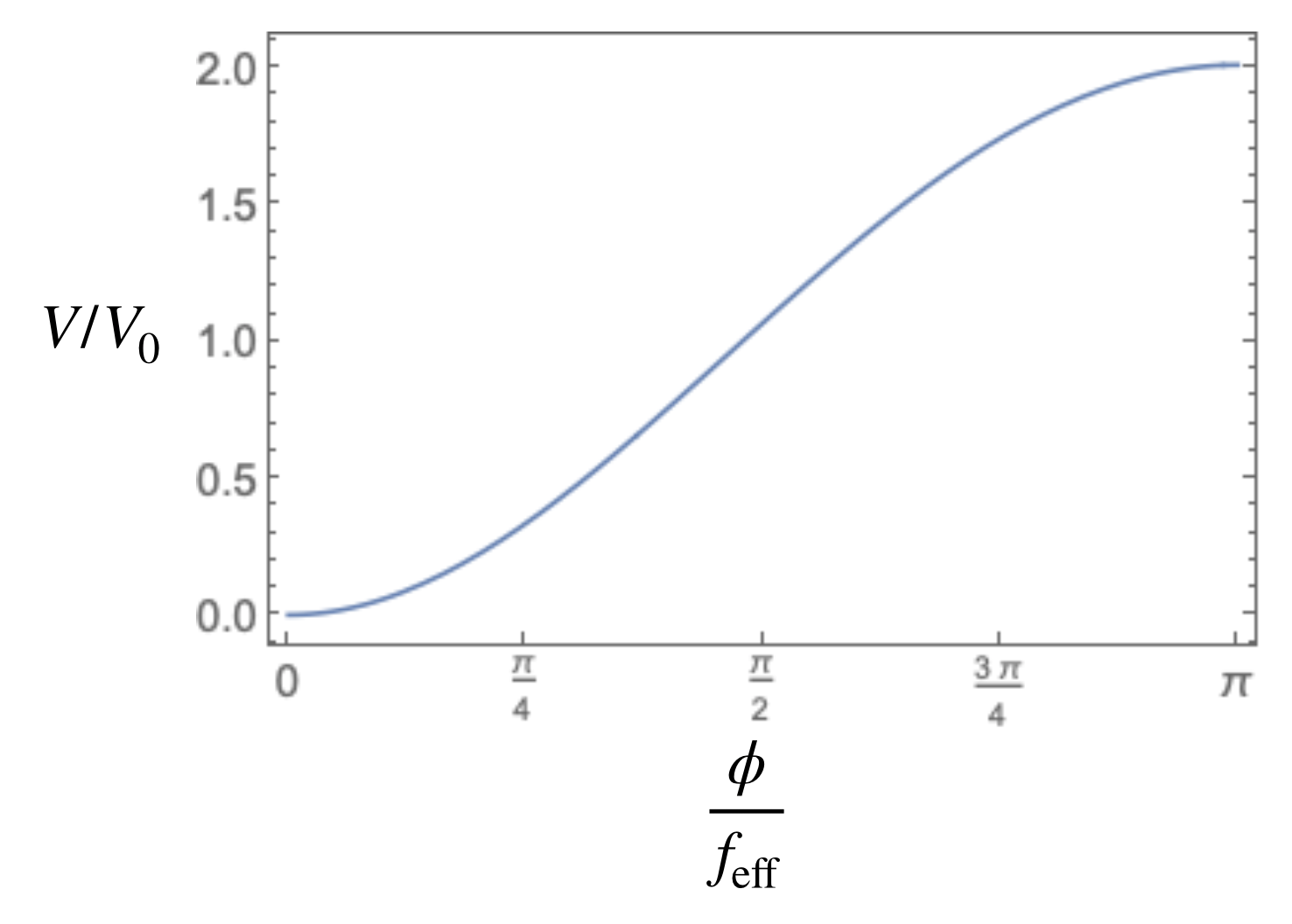}
% "\includegraphics" is very powerful; the graphicx package is already loaded
\caption{\label{fig:potential} The potential for the minimal gauge inflation model in $\phi/f_{\rm eff}=(0,\pi)$.}
\end{figure}

%=====================================================
\section{Minimal gauge inflation and the Conditions of refined Swampland conjecture}
\label{sec:model}
%=====================================================

A minimal gauge inflation model is based on a higher dimensional theory of non-Abelian ${\rm SU}(2)$ gauge symmetry on the orbifold, $S^2/\mathbb{Z}_2$~\cite{Gong:2018jer} with only a few free parameters, the compactification radius, $R$, and the gauge coupling constant, $g_5$ in five dimensions. The model is supposed to be the simplest realization of this category~\cite{ArkaniHamed:2003wu, Kaplan:2003aj,Park:2007sp, Kubo:2001zc}. The inflaton is identified with the extra dimensional component of the gauge field, $A_5\sim \phi$, and its potential is protected by the higher dimensional gauge symmetry itself but generated at one-loop level by gauge self interactions. As a result, the model is extremely predictive.\footnote{In Ref.~\cite{Cacciapaglia:2005da}, the electroweak symmetry breaking is realized by a fully radiatively generated potential.} It is desired to check if this model is consistent with the refined dS conjecture.

 The inflaton potential of minimal gauge inflation model is
\begin{eqnarray}
V(\phi) 
&=& V_0 \sum_{n=1}^\infty \frac{1}{n^5}\left[1-\cos \frac{n\phi}{f_{\rm eff}}\right] \\
&=&-\frac{V_0}{2} \left[{\rm Li}_5(e^{i\phi/f_{\rm eff}}) +{\rm Li}_5(e^{-i\phi/f_{\rm eff}}) -2 \zeta(5) \right],\nonumber 
\end{eqnarray}
where two of the most important model parameters, $V_0$ and $f_{\rm eff}$, are introduced: $V_0 = \frac{9}{(2\pi)^6 R^4}$ is  the scale of the potential  and $f_{\rm eff} =\frac{1}{\sqrt{2 \pi R}g_5} = \frac{1}{2\pi g_4 R}$ is the `effective decay constant'. 
%They are completely determined by the gauge coupling constant $g_5$ in $5$D and the compactification radius $R$. %We take these as two of the most important model parameters: $$(V_0,f_{\rm eff})$$ 
The potential is composed of infinitely many periodic terms. It has a maximum at $\phi/f_{\rm eff} =\pi$ and we only consider the physical region in $\phi/f_{\rm eff} =(0,\pi)$.  The inflation starts below $\phi/f_{\rm eff}\approx \pi$ then roll-down to the true vaccum at $\phi/f_{\rm eff}=0$. The model may be regarded as a UV completion of the natural inflation models~\cite{Freese:1990rb, Adams:1992bn,Kim:2004rp}.  The shape of the potential is depicted in Fig.~\ref{fig:potential}.

%%%%%%%%%%%%%
\subsection{Condition-1}
%%%%%%%%%%%%%

\begin{figure}[t]
\centering % \begin{center}/\end{center} takes some additional vertical space
\includegraphics[width=.95\columnwidth]{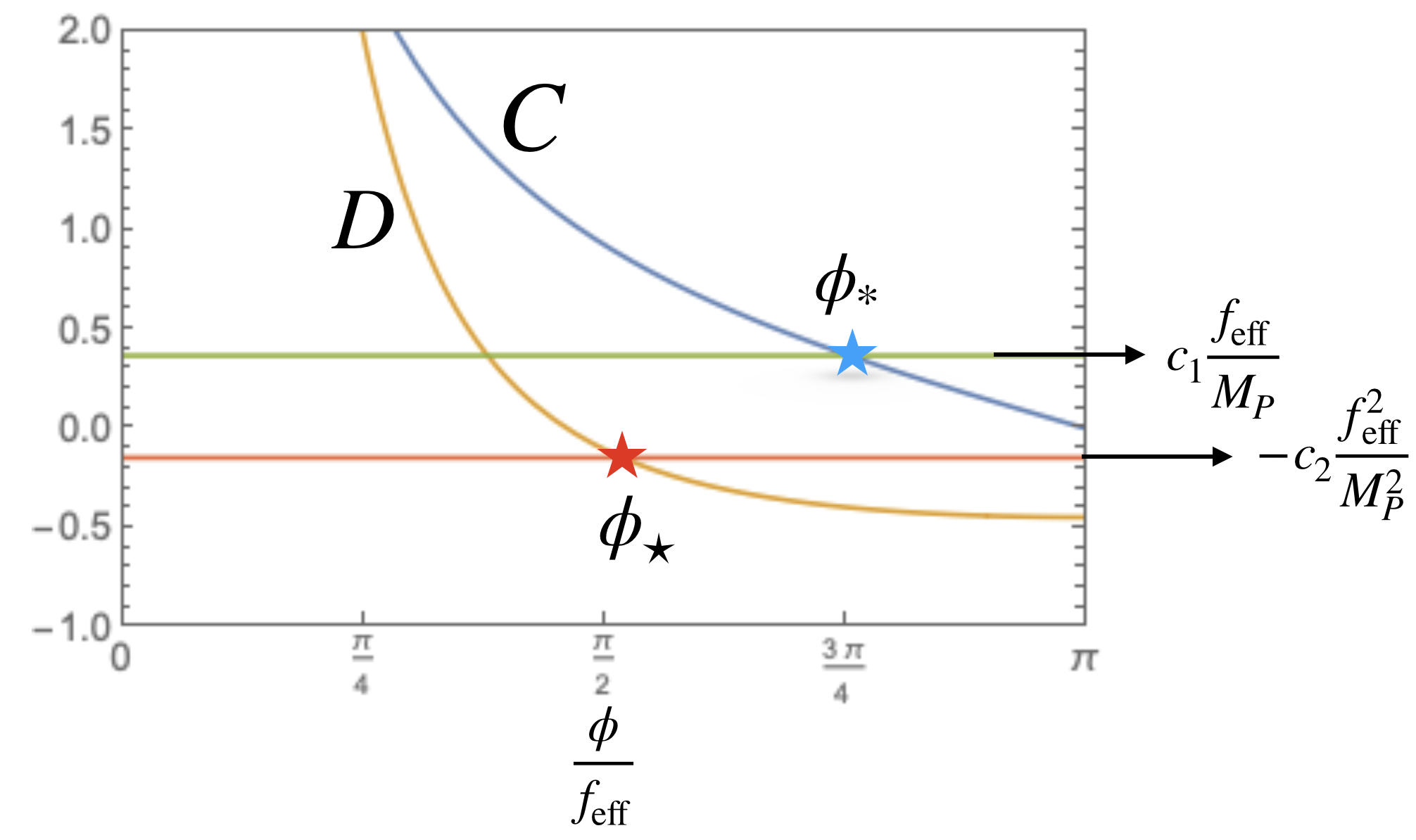}
% "\includegraphics" is very powerful; the graphicx package is already loaded
\caption{\label{fig:CD} $C(\phi/f_{\rm eff})$ and $D(\phi/f_{\rm eff})$ in $\phi/f_{\rm eff}=(0,\pi)$.
The locations of $\phi_*$ and $\phi_\star$ are also depicted.
}
\end{figure}

We first request (Condition-1) or $M_P ||V'||/V \geq c_1$. 
A convenient function is introduced:
\begin{align}
C\left(\frac{\phi}{f_{\rm eff}}\right) 
&\equiv f_{\rm eff} \frac{||V'||}{V} \\
&=\frac{i\left[{\rm Li}_4(e^{-i \phi/f_{\rm eff}})-{\rm Li}_4(e^{i \phi/f_{\rm eff}})\right]}{\left[{\rm Li}_5(e^{i\phi/f_{\rm eff}}) +{\rm Li}_5(e^{-i\phi/f_{\rm eff}}) -2 \zeta(5)\right]}.
\end{align} 
The function is plotted  in Fig.~\ref{fig:CD}~(upper curve). As the function is monotonically decreasing within the range of our interest, (Condition-1) sets up a upper limit for $\phi$: 
\begin{align}
&C\left(\frac{\phi}{f_{\rm eff}}\right) \geq c_1 \frac{ f_{\rm eff}}{M_P} \nonumber \\
 &\Leftrightarrow ~~\phi \leq \phi_* = f_{\rm eff}~ C^{-1}\left(c_1\frac{f_{\rm eff}}{M_P}\right).
 \label{eq:C1}
\end{align}

The critical value, $\phi_*$, is determined for a given value of $c_1 f_{\rm eff}/M_P$.
In principle, the parameter is constrained by the inflationary observables. We  will be discuss this in Sec.~\ref{sec:predictions}.

%

%%%%%%%%%%%%%
\subsection{Condition-2}
%%%%%%%%%%%%%

The potential is convex ($V''>0$) for a small field value of $\phi/f_{\rm eff}$ then becomes concave ($V''<0$) toward the plateau located at the top. The inflection point ($V''=0$) locates at $\phi/f_{\rm eff} \approx 1.45$. 
 To analyze the second condition, we introduce a convenient function, $D$, which encapsulates the information of the curvature of the potential:
\begin{align}
D\left(\frac{\phi}{f_{\rm eff}}\right) 
&\equiv f_{\rm eff}^2 \frac{V''}{V} \\
&=-\frac{{\rm Li}_3(e^{-i \phi/f_{\rm eff}})+{\rm Li}_3(e^{i \phi/f_{\rm eff}})}{\left[{\rm Li}_5(e^{i\phi/f_{\rm eff}}) +{\rm Li}_5(e^{-i\phi/f_{\rm eff}}) -2 \zeta(5)\right]}.
\end{align}
The shape of the $D$ is depicted in Fig.~\ref{fig:CD} (lower curve).

The function is monotonically decreasing and (Condition-2) limits the validity range of $\phi$ and sets the lower bound $\phi_\star$ as
\begin{align}
D\left(\frac{\phi}{f_{\rm eff}}\right) &\leq -c_2 \frac{f_{\rm eff}^2}{M_P^2} \nonumber \\
 \Leftrightarrow 
 ~~&\phi \geq \phi_\star = f_{\rm eff}~ D^{-1}\left(-c_2\frac{ f_{\rm eff}^2}{M_P^2}\right).
 \label{eq:C2}
\end{align}
%

%%%%%%%%%%%
\subsection{Condition-1 and Condition-2}
%%%%%%%%%%%

In principle, (Condition-1) in Eq.~\ref{eq:C1} and (Condition-2) in Eq.~\ref{eq:C2} are independent. 
However, at least one of the conditions can be satisfied in the whole region of $\phi$ if the lower bound of (Condition-2), $\phi_\star$, is smaller than the upper bound of (Condition-1), $\phi_*$ or 
\begin{eqnarray}
 \frac{\phi_\star}{f_{\rm eff}}=D^{-1}\left(-c_2 \frac{f_{\rm eff}^2}{M_P^2}\right) \leq C^{-1}\left(c_1\frac{f_{\rm eff}}{M_P}\right) =\frac{\phi_*}{f_{\rm eff}}.
\end{eqnarray}
This inequality is a generic condition that should be taken as the guiding principle of a model, which has similar features: growing potential having an inflection point in the middle toward the top of the potential.  There are many examples of this kind including the Higgs inflation~\cite{Bezrukov:2007ep, Bezrukov:2009db, Hamada:2014iga, Hamada:2014wna,Bezrukov:2014ipa,Bezrukov:2014bra}.

If the above condition is not satisfied, there exists a region of $\phi \in (\phi_*, \phi_\star)$ where neither condition is satisfied. Indeed, as one can notice in the figure, $C\gsim D$ in $\phi/f_{\rm eff} \gsim \pi/4$ so that one can actually find a set of values $(c_1,c_2)$ for a given value of $f_{\rm eff}/M_P$ satisfying the desired condition as one can clearly see in Fig.~\ref{fig:CD}.  For instance, with $(c_1,c_2) =(0.3,0.1)$ and $f_{\rm eff}/M_P\approx 1.2$, $(\phi_\star,\phi_*)\approx (1.6f_{\rm eff},2.4f_{\rm eff})$ thus the dS criterion is satisfied.

\begin{figure}[t]
\centering % \begin{center}/\end{center} takes some additional vertical space
\includegraphics[width=.95\columnwidth]{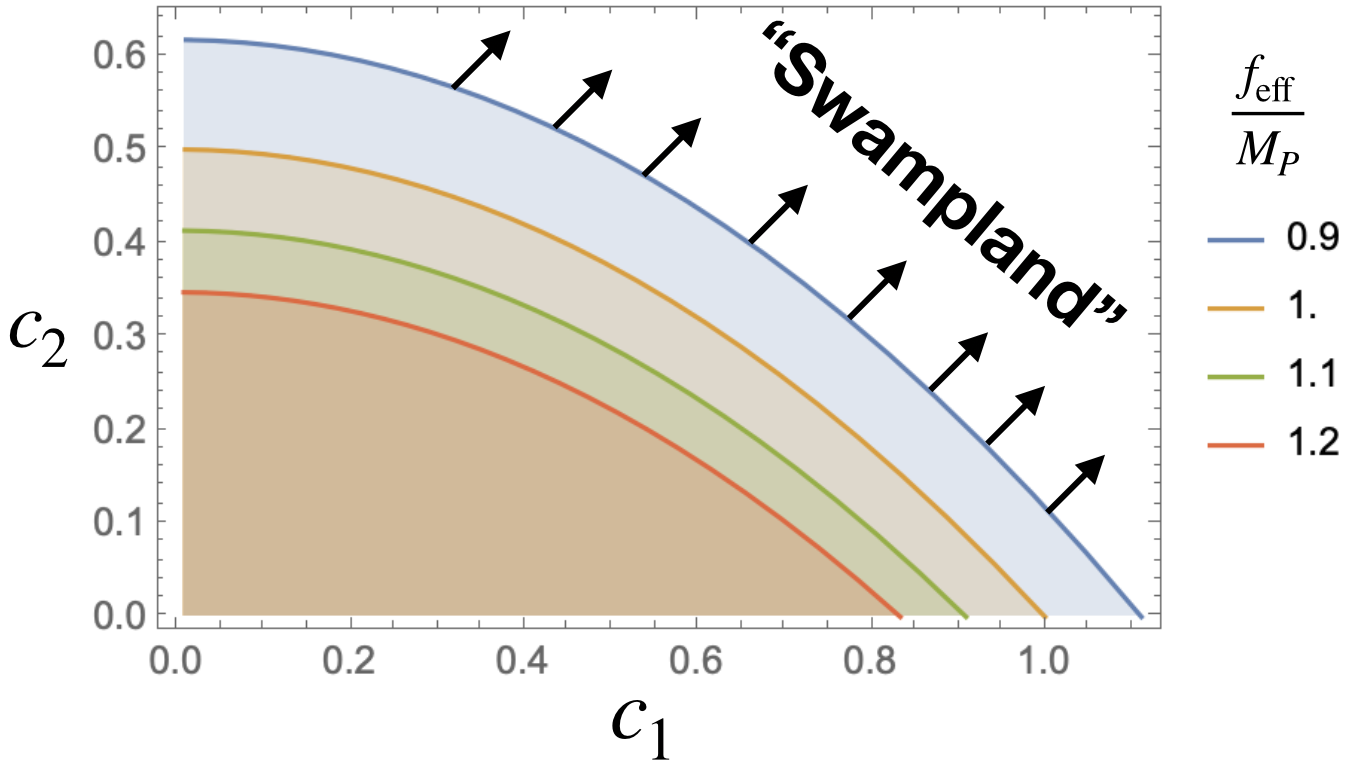}
% "\includegraphics" is very powerful; the graphicx package is already loaded
\caption{\label{fig:c1c2} The critical line of $(c_1,c_2)$. The region above the line is excluded by the dS conjecture or belongs to the ``Swampland".
}
\end{figure}

In Fig.~\ref{fig:c1c2} we depicted the critical lines in the plane of $(c_1,c_2)$ for various values of $f_{\rm eff}$:  $f_{\rm eff}/M_P =0.9, 1.0, 1.1, 1.2$. The regions below the lines (colored parts), the critical condition, $\phi_\star \leq \phi_*$, is satisfied thus the condition of the refined dS conjecture is fulfilled and belongs to the ``Landscape" but in the regions above the lines, the potential may not have a consistent quantum gravitational UV completion or belongs to the ``Swampland".  

Until this far, we examine the implications of the refined dS conjecture to generic field space of minimal gauge inflation model. Using the method, we found that the allowed range of $(c_1,c_2)$ depends on $f_{\rm eff}$. On the other hand,  if $(c_1,c_2)$ are known {\it a priori}, we can set the theoretically preferred range of the model parameters. Instead, in this paper, we take the observational data as the guideline of a theory and try to set the bound on $(c_1,c_2)$ in the next section. 
%Since $f_{\rm eff}$ is determined by the coupling constant $g_5$ and the size of extra dimension $R$, we may learn about the size of extra dimension. 

%\begin{widetext}
%
%

%\vspace{1.0cm}

%=====================================================
\section{Inflationary predictions}
\label{sec:predictions}
%=====================================================

%
\begin{figure}[t]
\centering
\includegraphics[width=.98\columnwidth]{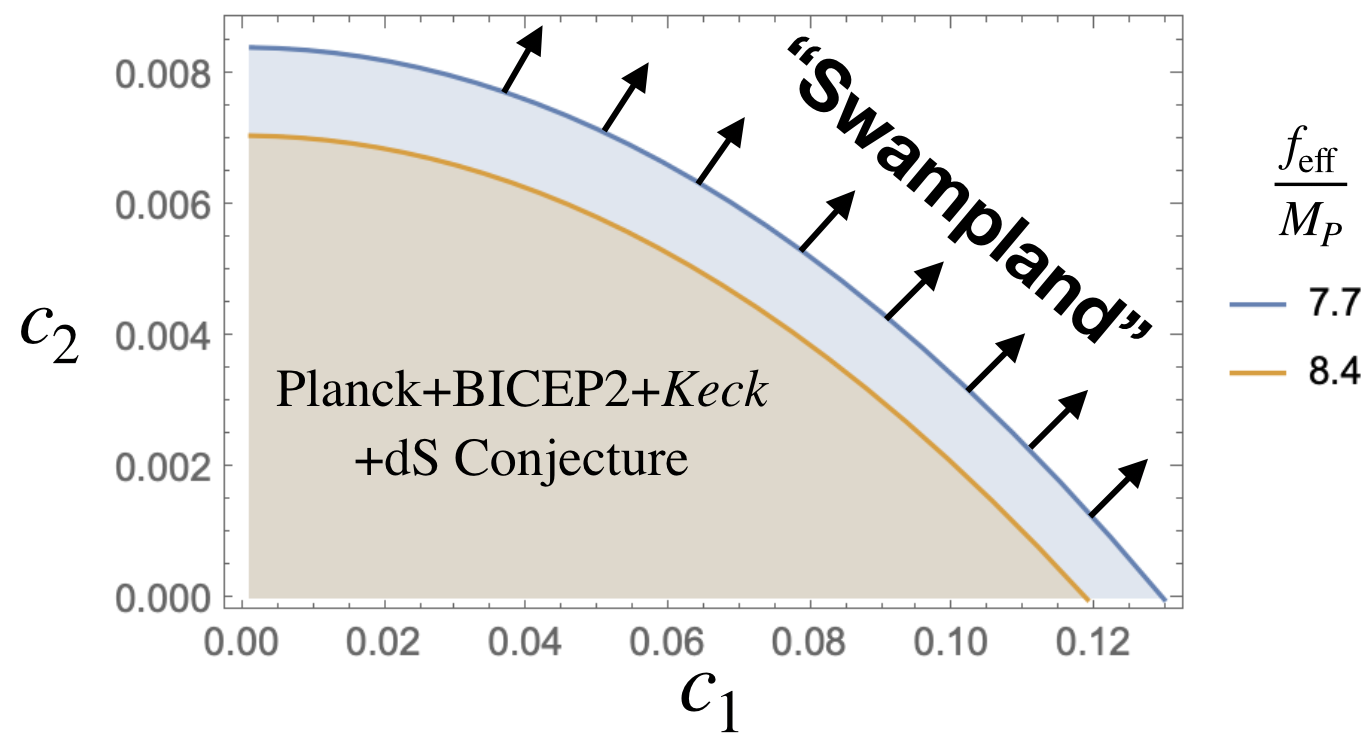}
% "\includegraphics" is very powerful; the graphicx package is already loaded
\caption{\label{fig:exp} The critical line of $(c_1,c_2)$ consistent with the observational data from Planck+BICEP2+{\it Keck}~\cite{Akrami:2018odb, Aghanim:2018eyx, Ade:2015xua,Ade:2018gkx}.
\label{fig:exp}}
\end{figure}

The inflationary observables, the power spectrum of the curvature perturbation, ${\mathcal P}_\zeta$, the corresponding spectral index, $n_s$, the tensor-to-scalar ratio, $r$,  of minimal gauge inflation are obtained in Ref.~\cite{Gong:2018jer}: 
 %, $e^{-N_e M_P^2/f_{\rm eff}^2}\ll 1$, 
\begin{align}
{\mathcal P}_\zeta
%&= \frac{1}{24\pi^2}\frac{V}{\epsilon M_P^4} \nonumber \\
%&=-\frac{V_0}{24\pi^2 M_P^4} \left(\frac{f_{\rm eff}}{M_P}\right)^2 \frac{{\rm Li}_5(e^{-i\phi/f_{\rm eff}})+{\rm Li}_5(e^{i\phi/f_{\rm eff}})-\zeta(5)}{C\left(\frac{\phi}{f_{\rm eff}}\right)^2} \nonumber \\
&\approx \frac{V_0 f_{\rm eff}^2}{6\pi^2 M_P^6}e^{-N_e M_P^2/f_{\rm eff}^2},
\\
n_s
%&= 1 - 6\epsilon + 2\eta \nonumber \\
%&= 1-3 \left(\frac{f_{\rm eff}}{M_P}\right)^{-2}C\left(\frac{\phi}{f_{\rm eff}}\right)^2 -2\left(\frac{f_{\rm eff}}{M_P}\right)^{-2}D\left(\frac{\phi}{f_{\rm eff}}\right)^2   \nonumber \\
&\approx 1-\frac{M_P^2}{f_{\rm eff}^2},  \\%\left(1+2e^{-N_e M_P^2/f_{\rm eff}^2}\right), \\
%~~~~
r 
%&= 16\epsilon \nonumber \\
%&= 8 \left(\frac{f_{\rm eff}}{M_P}\right)^{-2}C\left(\frac{\phi}{f_{\rm eff}}\right)^2 \nonumber \\
&\approx 8\frac{M_P^2}{f_{\rm eff}^2} e^{-N_e M_P^2/f_{\rm eff}^2}, % \left(1+e^{-N_e M_P^2/f_{\rm eff}^2}\right),
\end{align}
%\end{widetext}
where $N_e$ is the number of efolds. From these, we get  the model parameters and $N_e$ as
\begin{eqnarray}
V_0&\approx& \frac{48 \pi^2 {\mathcal P}_\zeta (1-n_s)^2}{r} M_P^4,\\
f_{\rm eff}&\approx& \frac{1}{\sqrt{1-n_s}} M_P,\\
N_e&\approx& \frac{1}{1-n_s} \log \frac{8(1-n_s)}{r}.
\end{eqnarray}
%

%The adjustable model parameters are $(V_0,f_{\rm eff})$ which are constrained by three observables $({\mathcal P}_\zeta, n_s, r)$  with the number of efolds, $N_e=50-60$.
%The model parameters are almost completely determined by observational data taking the reference values 
%
We take the reference values
\begin{align}
{\mathcal P}_\zeta =2.5 \times 10^{-9}, n_s =0.965 \pm 0.004,%~~r <0.06,
\end{align}
from the recent data of Cosmic Microwave Background radiation (CMB) from the Planck observatory~\cite{Akrami:2018odb, Aghanim:2018eyx, Ade:2015xua} and also the data taken by the BICEP2/{\it Keck} CMB polarization experiments~\cite{Ade:2018gkx} then we determine the model parameters, $V_0$ and $f_{\rm eff}$, as well as the allowed window for $r$ as
\begin{align}
\left.V_0\right|_{\rm Planck+BICEP2+{\it Keck}} 
&\approx (3.0-4.2)\times 10^{-8} M_P^4, \\
\left.f_{\rm eff}\right|_{\rm Planck+BICEP2+{\it Keck}} 
&\approx 5.3 M_P, \\
\left.r\right|_{\rm Planck+BICEP2+{\it Keck}}
&\approx 0.034-0.049,
\end{align}
for the requested number of efolds, $N_e=50-60$.  Notice that the tensor-to-scalar ratio $r$ is within the future probe. 
%Now the question is if there exists a good parameter choice, which is consistent with 
%Here we take the values of scalar spectral index ($n_s$) and tensor-to-scalar ratio ($r=T/S$)
The determined values of $V_0$ and $f_{\rm eff}$ would give the compactification radius and the gauge coupling constant:
\begin{align}
RM_P &\approx 7.7-8.4,  \nonumber \\
g_4 &\approx (3.5-3.9)\times 10^{-3},
\end{align}
which look  consistent with the quantum gravity and the perturbative gauge theory with $g_4 \ll 4\pi$.  %Allowing the experimental uncertainties, one may find the region of the parameter space in detail~\cite{Gong:2018jer}. 

Finally, having determined the input parameters of the model we now can directly check the dS conjecture. 
The Fig.~\ref{fig:exp} is depicted to show the parametric region of $(c_1,c_2)$ which is consistent with the dS conjecture as well as the observational data. %Compared to the results in Fig.~\ref{fig:c1c2}, 
The allowed values of $c_1$ and $c_2$ are typically $c_1\sim 0.15$ and $c_2 \sim 0.01$ or smaller.  The values are not strictly ${\cal O}(1)$ as requested in the conjecture but still close numerically.

%\vspace{1.0cm}

%=====================================================
\section{Conclusion}
\label{sec:conclusion}
%=====================================================

The latest swampland conjecture could provide important implications to the low energy effective theory models which may or may not be consistent with the quantum gravity theory. The conjecture suggests two related but independent conditions for  any scalar potential $V(\phi)$ of a low energy effective theory of a consistent  quantum gravity:
\begin{align*}
||\nabla V|| \geq c_1\frac{V}{M_P},~~~ {\rm min}(\nabla_i \nabla_j V) \leq -c_2\frac{V}{M_P^2},
\end{align*}
which we call (Condition-1) and (Condition-2), respectively, in this paper. The parameters $c_1$ and $c_2$ are supposed to be universal but unknown positive constants.  In this paper, we closely examine a minimal gauge inflation model as a concrete example of potentially realistic inflationary model and apply the dS conjecture to see the consistency with a quantum gravity theory. Interestingly, the potential indeed allows a parametric region in $(c_1 \lsim 1,c_2 \lsim 1)$.  If we apply the latest cosmological observations from Planck 2018 and also BICEP2+{\it Keck}, the allowed region shrinks but still exists as is clearly seen in Fig.~\ref{fig:exp}. Finally, we would emphasize that  the method developed in this paper can be applied to {\it any theory} with a similarly behaving potential: growing monotonically, having an inflection point on the way to the top.

%======================================================================
%{\bf Acknowledgements:} 
\acknowledgments
SC is thankful to Kohei Kamada  for valuable comments and Matt Reece for discussion during the CERN-TH institute in summer 2018. This work was supported in part by the National Research Foundation of Korea (NRF) grant funded by the Korean government (MSIP) (No.2016R1A2B2016112) and (NRF-2018R1A4A1025334).

\bibliography{mgibib}
\bibliographystyle{apsrev4-1}
%\nocite{*}

\end{document}